\documentclass{article} % For LaTeX2e
\usepackage{iclr2017_conference,times}
\usepackage{hyperref}
\usepackage{graphicx}
\usepackage{pythonhighlight}
\usepackage[noend]{algpseudocode}
\usepackage{algorithmicx,algorithm}
\usepackage{url}

\title{Dragon: A Computation Graph Virtual Machine Based Deep Learning Framework}

\author{Ting Pan  \\
SeetaTech, Beijing, 100049, China \\
University of Chinese Academy of Sciences, Beijing, 100049, China \\
\texttt{ting.pan@seetatech.com}
}

\begin{document}

\maketitle

\begin{abstract}
Deep Learning has made a great progress for these years. However,
it is still difficult to master the implement of various models because
different researchers may release their code based on different frameworks or interfaces.
In this paper, we proposed a computation graph based framework which only aims to introduce well-known interfaces.
It will help a lot when reproducing a newly model or transplanting models that were implemented by other frameworks.
Additionally, we implement numerous recent models covering both Computer Vision and Nature
Language Processing. We demonstrate that our framework will not suffer from model-starving because
it is much easier to make full use of the works that are already done. Code will be available soon at \href{https://github.com/neopenx/Dragon}{https://github.com/neopenx/Dragon}.
\end{abstract}

\section{Introduction}

Deep Learning(\citet{Lecun2015}) models now have evolved much quicker than any time before.
Programming specifically for each model is not realistic, because massive common modules can be re-used.
Thus a lot of works(\citet{Theano}, \citet{Caffe}, \citet{MXNet}, \citet{TensorFlow}) have being done to support various deep neural nets and related applications.

As more and more frameworks have been developed, it towards a problem that is hard learn them for users,
especially memorizing some strange interfaces without a handbook. Generally speaking, those architectures look all same
if you divide them into two parts:

\begin{itemize}
\item Frontend(mainly including net definitions and arguments)
\item Backend(mainly including efficient implement of operators)
\end{itemize}

Diverse frameworks often carry similar backends but disparate frontends, which contributes an idea for designing a new framework
to support various frontends but share a common backend. It is not rare in the field of compiling since a decade ago, as is well known,
LLVM(Low Level Virtual Machine) project build a environment for several programming languages. Inspired by this, we introduce
a virtual machine between frontend and backend, disruption is avoided when mixing multiple frameworks' interfaces even in one source file.

Our work early comes from these computation graph based framework: Caffe2, MXNet, and TensorFlow,
but we strive to expose computation graph as a public interface without any complicated encapsulation,
leading simpler and cheaper debugging. After rethinking the successful elements in Caffe, we found that
both layer-specific direct programming and script-specific direct defining make it so popular.
By exposing the computation graph directly, operators and their arguments are clear before executing,
which provides the same user experience as Caffe.

\section{Computation Graph}

\subsection{Function as A Graph}

\citet{Bengio2009} provide a general perspective for common functions, as shown in Fig.1:
\begin{figure}[t]
  \centering
  \includegraphics[height=0.5\textwidth]{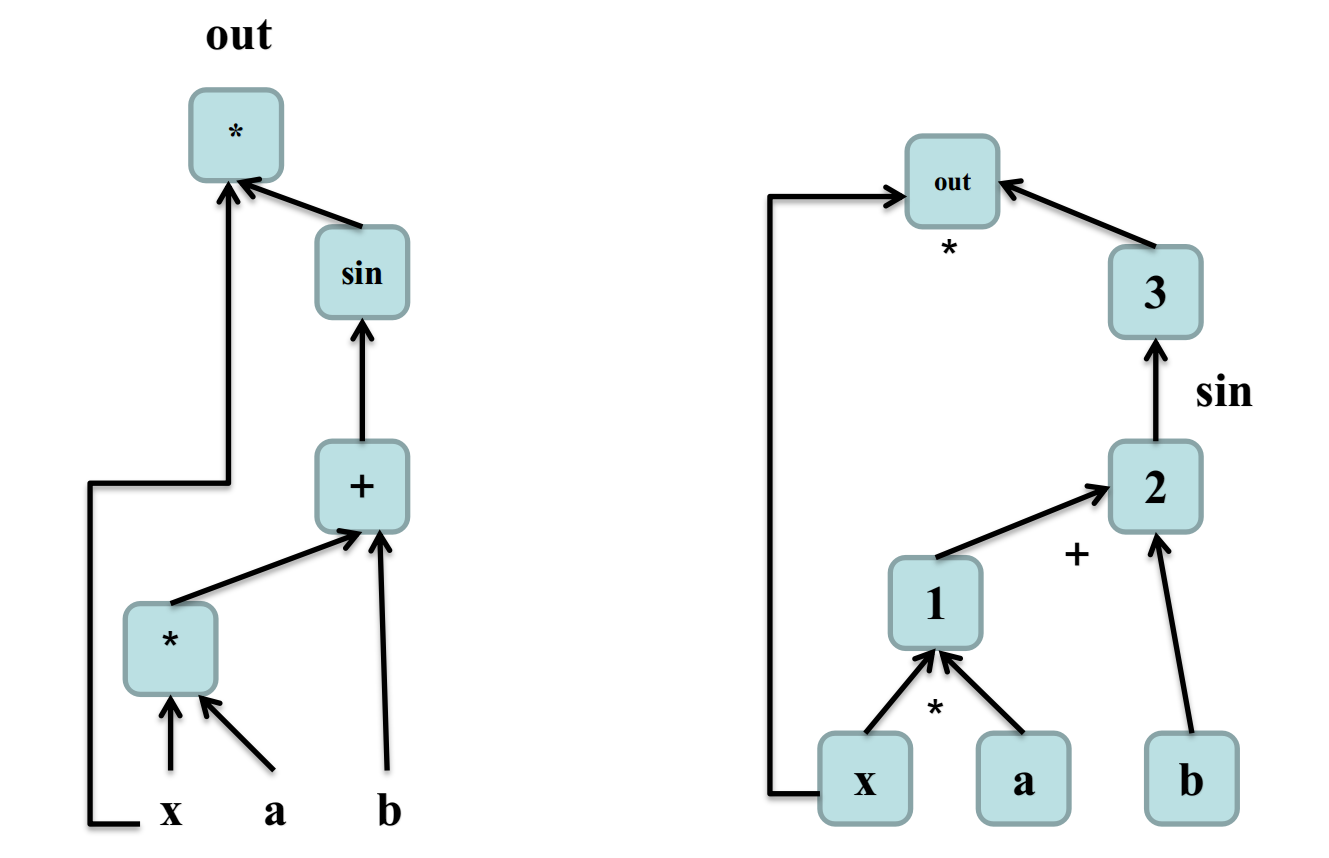}
  \caption{Two different approaches to utilize a graph, Left(\citet{Bengio2009}), Right(Ours).
  We prefer right for simplicity during traversing a graph. See more in Sec 2.2.}
\end{figure}

A function can be expressed as the composition of operators, intermediate results, inputs and outputs.
Considering we almost perform linear algebra operations, tensors(n-dimension array) are used to replace latter for simplicity.
If we regard these tensors as nodes, which converts into a problem finding all available paths from inputs to outputs,
operators will be selected to establish connections through these paths.

A similar abstraction was described by TensorFlow(\citet{TensorFlow}), but we use nodes to represent tensors but not operators.
It is convenient to connect tensors into a DAG(Directed Acyclic Graph) for coloring, as done in Caffe(\citet{Caffe}),
whose blobs are bottom-top associated. An un-colored tensor should not be solved,
and a operator with all outputs un-colored should be removed.

Instead of setting them manually, (\citet{Theano} provide a simulated "function" to collect targets:
\begin{center} \emph{f \ = \ function(inputs, \ outputs)} \end{center}
It means that if a function has already been defined, push specific inputs will get desired outputs.
Different from traditional imperative programs, declarative programs emphasize to build a re-used model,
which is more appropriate for machine learning.

\subsection{Graph Optimization: Forward Prune}

Assume that the set of nodes is $ X  = \{ x_{1...n} \}$, A pair of nodes $(x_{i}$, $x_{j})$
can be connected by a directed edge $x_{i} \rightarrow x_{j}$ if existing a operator takes $x_{i}$ as one
of inputs and $x_{j}$ as one of outputs.

Let us define a function:

\begin{center} $f \leftarrow xsin(ax + b) $ \end{center}

A common idea is to choose nodes whose in-degree is zero as starting points,
traverse all adjoint and connected nodes sequentially, which can be computed in $O(n)$ if marking those visited.

However, there may be such nodes that can not affect solving targets, consider we just want to solve:

\begin{center} $ y = ax $ \end{center}

then, $b$ is redundant and should not be marked.

It is probably easier to make an analogy between deriving a formula and running a neural network.
Backward pass from targets will make a equivalent traversing but can ignore irrelevant inputs. \\
Algorithm.1 shows how to prune redundant inputs through a backward traversing in $O(n)$.

\begin{algorithm}[t]
\caption{Backward Traversing for Forward-Pass}
\hspace*{0.02in} {\bf Input:}
\hspace*{0.02in} Graph: $ G = \{ x_{1...n} \} $ \\
\hspace*{0.5in} Solving targets: $ T = \{ t_{x_{1}...x_{m}} \} $ \\
\hspace*{0.02in} {\bf Output:}
Marked flags: $ M = \{ m_{x_{1}...x_{n}} \} $ \\
\hspace*{0.02in} {\bf Define:} DepthFirstSearch($x$)
\begin{algorithmic}[]
\State $m_{x} \leftarrow$ True
    \For {$x_{parent}$ in $Parent_{x}^{G} $}
        \If {not $m_{x_{parent}}$}
            \State DepthFirstSearch($x_{parent}$)
        \EndIf
    \EndFor
\end{algorithmic}
\begin{algorithmic}[1]
\For{$t_{\hat{x}}$ in $T$}
　　\If{not $m_{\hat{x}}$}
        \State DepthFirstSearch($t_{\hat{x}}$)
    \EndIf
\EndFor
\end{algorithmic}
\end{algorithm}

\subsection{Graph Optimization: Backward Prune}

Automatic Differentiation(\citet{Griewank2000}) now has became a widespread tool in graph based deep learning frameworks.
It is not a magic but a trick to fake users by wrapping the classic Back-Propagation(\citet{Hinton1986}) interfaces.

\citet{Caffe} show that: if only a layer provide a function to compute element-wise local gradients,
then a global net can simulate the chain rule by executing these functions from top to bottom.

However, simply providing backward process for each layer will also lead redundant computation,
as we often require gradients with respect to several specific inputs(e.g. weights and bias).

Thus, if we unfold all backward processes at the end of last forward process,
then it turns into a two-stage DAG. By performing the forward traversing from a objective target to a derivative target,
one chain(path) is built for executing. Repeat this step until enumerating for all pairs $ (x_{obj}, x_{wrt}) $,
we can eliminate all $ x_{nomark} $.

The problem can be also regarded as to search a optimal substructure for all nodes,
if we re-define the number of maintaining dependent edges as cost.
Perform a naive search will leading a exponential cost, by leveraging dynamic programming, we can solve it in $O(n)$.
Algorithm.2 shows how to perform a memorized-search through a forward traversing.

\begin{algorithm}[t]
\caption{Forward Traversing for Backward-Pass}
\hspace*{0.02in} {\bf Input:}
\hspace*{0.02in} Graph: $ G = \{ x_{1...n} \} $ \\
\hspace*{0.5in}  Pairs: $ P = \{(x_{obj}^{1}, x_{wrt}^{1})...(x_{obj}^{m}, x_{wrt}^{m})\} $ \\
\hspace*{0.5in}  Visited flags: $ V = \{ v_{x_{1}...x_{n}} \} $ \\
\hspace*{0.02in} {\bf Output:}
Marked flags: $ M = \{ m_{x_{1}...x_{n}} \} $ \\
\hspace*{0.02in} {\bf Define:} DepthFirstSearch($x$, \, $x_{wrt}$, \, $Path$)
\begin{algorithmic}[]
\If {$v_{x}$ \textgreater \, 0} \quad \quad { // visited already }
    \If {$v_{x}$ is 2} \quad \quad { // connectived from $x$ to $x_{wrt}$ }
        \For{$\hat{x}$ in $Path$}
            \State $ v_{\hat{x}} \leftarrow 2 $
            \State $ m_{\hat{x}} \leftarrow $ True
        \EndFor
    \Else \State return
    \EndIf
\EndIf
\State $v_{x} \leftarrow $ 1
\For {$x_{child}$ in $Child_{x}^{G} $}
    \State $ NewPath  \leftarrow  Path $
    \State $ NewPath.append (x_{child})$
        \If {$x_{child}$ is $x_{wrt}$}  \quad \quad { // connectived from $x_{obj}$ to $x_{child}$ }
            \For{$\hat{x}$ in $ NewPath $}
                \State $ v_{\hat{x}} \leftarrow 2 $
                \State $ m_{\hat{x}} \leftarrow $ True
            \EndFor
            \State return
        \EndIf
    \State DepthFirstSearch($x_{child}$, \, $x_{wrt}$, \, $NewPath$)
\EndFor
\end{algorithmic}
\begin{algorithmic}[1]
\For{$(\hat{x_{obj}}, \hat{x_{wrt}})$ in $P$}
　　\State Clear(V)
    \State DepthFirstSearch($\hat{x_{obj}}$, \, $\hat{x_{wrt}}$, \, $\{x_{obj}\}$)
\EndFor
\end{algorithmic}
\end{algorithm}

\subsection{Graph Optimization: In-Place}

Operators whose inputs will not be used in future can share memory for outputs,
most common non-linear functions support this specific, such as Sigmoid, Tanh, ReLU(\citet{Nair2010} \& \citet{Glorot2011}) and Dropout(\citet{Srivastava2014}.

Different from \citet{Caffe}, we adopt a adaptive in-place policy based on graph,
by accepting a basic input-output definition, it returns a optimized graph, Fig.2. showing one structure for this.

\begin{figure}[t]
  \centering
  \includegraphics[height=0.4\textwidth]{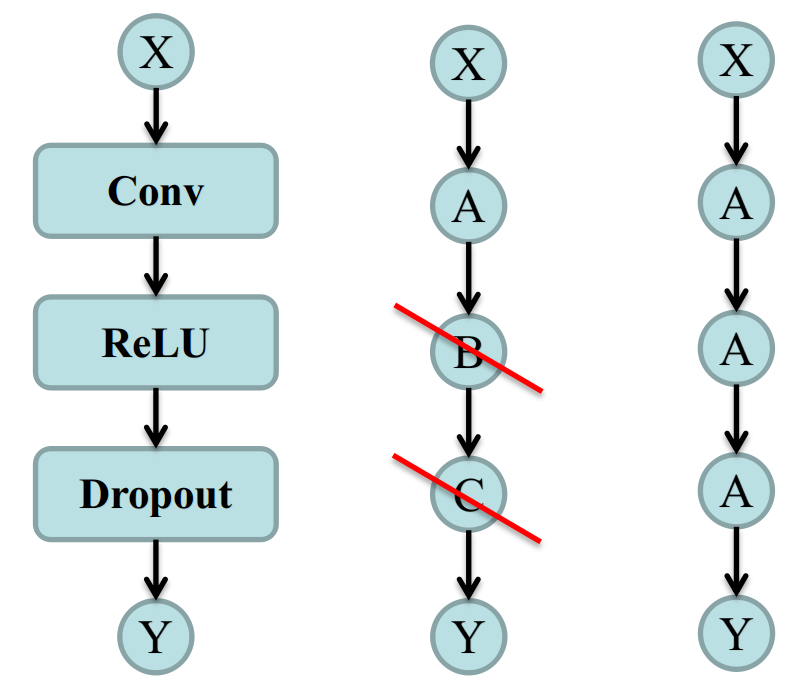}
  \caption{A structure used in deep neural networks, Left(structure), Middle(basic graph), Right(optimized graph).
  Red oblique line represents this node can be shared.}
\end{figure}

We generalize a rule that the in-place structures should contains nodes which only have at most one child,
then it turns into a problem to find a longest path according with above when giving a ancestor, and all nodes on the path
can rename as ancestor. Note that rename nodes through a path will break the graph into a cyclic graph, which should be done after graph prune.

Algorithm.3 shows how to determine all available in-place structures in $O(n)$.

\begin{algorithm}[t]
\caption{Forward Traversing for Available In-place Structures}
\hspace*{0.02in} {\bf Input:}
\hspace*{0.02in} Graph: $ G = \{ x_{1...n} \} $ \\
\hspace*{0.02in} {\bf Output:}
Rename Dictionary: $ RD = \{ \} $ \\
\hspace*{0.02in} {\bf Define:} DepthFirstSearch($x$, $ancestor$)
\begin{algorithmic}[]
\If {$x$ in $RD$}
    \State Return
\EndIf
\State $RD[x] \leftarrow ancestor$
\If {len($Child_{x}^{G}$) is 1}
    \State DepthFirstSearch($x_{child}$, $ancestor$)
\EndIf
\end{algorithmic}
\begin{algorithmic}[1]
\For{$\hat{x}$ in $G$}
    \State DepthFirstSearch($\hat{x}$, $\hat{x}$)
\EndFor
\end{algorithmic}
\end{algorithm}

\section{Architectures}

In this section, we will describe the architectures designed in Dragon.
Dragon takes a compact backend comparing to other graph based frameworks in order to support virtual machine,
we introduce four components: \emph{Tensor}, \emph{Operator}, \emph{Graph} and \emph{Workspace}. Dependencies were carefully considered and only protobuf is necessary.

\subsection{Workspace}

Traditional frameworks allocate memory inside models,
which often draws a critical memory re-allocating if models are relevant.
We investigate such models like: Cross-Validation(\citet{Bishop2006}), DCNNs(\citet{Krizhevsky2012}) and GANs(\citet{Goodfellow2014}),
which will reduce a large amount memory if sharing those allocated.

We follow the design of Caffe2 by introducing a \emph{Workspace} to take charge of all graphs and tensors,
while tensors are moved out from graphs(i.e. a tensor is equal to a graph in the workspace).
Each tensor has a unique name and can be fetched or feeded in any graphs, operators or even at the frontend,
which provides convenience for memory sharing and re-using.

Setting up a workspace leading architectures into a modern design pattern,
which is well-known as MVC(Model-View-Controller).
We found it essential for most machine learning applications by leveraging devices more efficiently,
Table 1. shows a memory-used comparison between workspace based kernel and cuDNN(\citet{CUDNN}) kernel on ResNet-50(\citet{He2015}).

\begin{table}[t]
\caption{Memory-used Comparison on ResNet-50(\citet{He2015})}
\label{sample-table}
\begin{center}
\begin{tabular}{ccc}
\multicolumn{1}{c}{\bf Batchsize} & \multicolumn{1}{c}{\bf Workspace}  & \multicolumn{1}{c}{\bf cuDNN}
\\ \hline \\
4         &1.5Gb     &1.6Gb   \\
8         &2.5Gb     &2.6Gb       \\
16        &4.5Gb     &4.6Gb    \\
25        &6.8Gb     &6.8Gb \\
\end{tabular}
\end{center}
\end{table}

\subsection{Graph}

\emph{Graph} plays a key role in organizing the topological order when giving a optimized graph definition,
we strictly keep the creating order for available operators while a inverse order
for auto generated gradient operators to simulate Back-Propagation(\citet{Hinton1986}).

Un-used operators will be removed directly, but tensors not for the completeness of a connected-graph.
By simply renaming all un-used tensors as \emph{ignore} in the workspace,
we can avoid duplicated memory allocating even if they are computed.

Another important structure in Deep Learning(\citet{Lecun2015}) is updating weights sequentially.
As giving a pair of weight and gradient $ (W, g) $, \, require:

\begin{center} $ W = W - \alpha \cdot g $ \, where $\alpha$  represents learning rate  \end{center}

which also leads into a cyclic graph. To resolve this problem, we introduce a Updater at the frontend for collecting
all pairs of weight and gradient and other hyper-parameters(e.g. learning rate multiplier or weight decay),
then generate all required update operators such as
Momentum(\citet{Lecun1998}), RMSProp(\citet{Tieleman2012}) and Adam(\citet{Kingma2014}) into an independent graph,
which is similar to \emph{Solver} introduced by \citet{Caffe}.
The separation of computing and updating provides flexibility for supporting some coarse-grained frameworks or distributed training.

\subsection{Operator}

\emph{Operator} takes inputs and outputs through names from workspace, which manages memory independently, as done in Caffe2.

\emph{Layer}(\citet{Caffe}) now is split into dual operators, one for \emph{Run} and another for \emph{Gradient},
which brings a problem that how to utilize data computed in Run-Op but required in Gradient-Op.
We introduce an \emph{Anchor} property to bind them together: each Run-Op has a unique name for anchor while corresponding Gradient-Op shares
this anchor to fetch specific tensors from workspace.
We demonstrate it useful for programming arbitrarily complex operators. Moveover,
removing specific anchor will simply make tensor itself global shared, which is frequently cast during memory optimization in Dragon.

\subsection{Tensor}

\emph{Tensor} consists of name, shape, type and context memory. We highly integrate them together for simplicity.
TMP(Template Meta Programming) is used for various types and contexts, which aims to contribute a cross-platform kernel.

While supporting multiple contexts at the same time often leads to confused memory management,
we follow the \citet{Caffe}, and still use a State-Machine but bracing more contexts(e.g. OpenCL), called \emph{MixedMemory}.
\emph{MixedMemory} provides a safe and efficient memory swapping during context switching, that is helpful for eliminating common memory-misused bugs.

\section{Virtual Machine}

Although architectures above could contribute a tiny framework, we still desire it to be more humanized.
An interesting idea is that: basic interfaces of Theano can be used for simulating Caffe or TensorFlow,
thus Theano, Caffe and TensorFlow can share a common backend if providing enough kernels of operator.

In this section, we demonstrate a cross-frameworks frontend is feasible.
When any of participating crucial interfaces is not reasonable, we can use the alternatives.

\subsection{Theano}

We begin with Theano, an inception drifts toward a modern deep learning framework,
and prove that the following fundamentals are quite useful for composing extensive modules.

\begin{itemize}
\item theano.function

\emph{Function}(\citet{Theano}) could be made when only giving specific outputs,
which is an appropriate wrapper for running a \emph{Graph} that described in Sec 3.2.
However, it seems difficult to exact the definition of a graph from several outputs,
except that all expressions were recorded before.
For this reason, we also introduce \emph{Tensor} at the frontend to record history operators:
a new operator will be pushed back into a sequence where stores operators created before,
while creating order is recorded at the same time.
By parsing, sorting and hashing all operators stored in outputs,
we can automatically construct the definition of a graph,
which is more convenient than adding desired operators sequentially introduced by Caffe2
or editing a text one manually introduced by Caffe.

\item theano.tensor.grad

\emph{Grad} accepts an objective with respect to its parameters, perhaps it is the earliest tool for
\emph{Automatic Differentiation}(\citet{Griewank2000}) in Deep Learning.
As described in Sec 2.3, this could be directly implemented for collecting all derivative pairs,
then attach them to the definition of a graph for further graph optimizations.

\item theano.scan

\emph{Scan} supports a dynamic loop architecture while wraps any custom operators,
that lightens a lot for building \emph{Gibbs Sampling}(\citet{Geman1984}) or \emph{Recurrent Neural Networks}(\citet{Jordan1986}, \citet{Elman1990}, \citet{Hochreiter1997}, \citet{Cho2014}), especially when modeling a sentence with uncertain length.

It urges us to process dynamic graphs at the backend while still accept custom templates at the frontend.
We try to unfold a custom sub-graph based on current inputs during the runtime of the kernel,
it will not be costly if repeating the graph optimizations described in Section2 for multiple times.
Persist a graph with fixed length may also help accelerating computation because the steps of a loop are often limited.

\end{itemize}

\subsection{Tiny Dragon}

Combining with above, we get a plain wrapper around the backend, dubbed Tiny-Dragon,
which takes the light weight programming across tensor, operators and workspace. The basic codes as shown below:

\inputpython{code1.py}{1}{24}

Our goal is to reduce extra unnecessary structures or interfaces.
Therefore, in addition to feed or fetch, the last thing is designing a desirable function through available operators.
As more operators developed to cover both coarse-gained and fine-gained applications,
this tiny framework provides a fast, efficient and powerful way to model complex mathematic problems.

\subsection{Caffe}

Based on Tiny-Dragon, we fuse Caffe for coarse-gained applications. As a blob can be represented by dual tensors,
one for \emph{data} and another for \emph{diff}, transition from the fine-level to coarse-level between these two frameworks can be quite smooth by exploiting the merits
of essential interfaces below.

\begin{itemize}

\item caffe.Filler

Different from Theano or TensorFlow, Caffe do not enforce setting explicit shape for learnable parameters,
but relies on smart shape-inference and a filler with specific type,
that helps a lot for stacking very deep neural networks(\citet{Simonyan2014}, \citet{Szegedy2014}, \citet{He2015}).

We support both shape-inference filling and shape-fixed filling.
However, setting a mistaken shape through feeding will be checked by the following inference before executing.

\item caffe.PythonLayer

\emph{PythonLayer} bridges between frontend and backend, where all installed python packages or existing variables can be called for computing.
It was common used for data buffering(\citet{Shelhamer2017}) and complicated linear algebra computation(\citet{Girshick2015}, \citet{Ren2015}).
Caffe utilizes it by binding a C++ structure to a Python structure through \emph{Boost} library, which carries directly memory access for the frontend.
However, \emph{Boost} is so heavy that must be completely removed. We prefer an alternative to fetch or feed memory through workspace described in Sec 3.1.

\item caffe.Net

\emph{Net} provides a vivid abstraction of several end-to-end functional layers defined as text format.
As almost layers can be implemented equivalently by one or more operators, we absorb layer parameters while translating as operator arguments,
which makes it quick to train or deploy caffemodels.

Instead of placing the abstraction of \emph{Net} into the backend, we strive to reproduce it through \emph{Function} implemented above.
By collecting input/output blobs and learning rate multipliers, \emph{Function} could be an alternative to \emph{Net}.
Explicit forward or backward process can be also easily implemented through filtering depends on the type of operators.

Considering manual In-place occurs frequently in the text-format definitions,
we try to allocate unique tensors for top blobs at the frontend,
while perform optimization together with other frameworks at the backend,
which is more significant than those frameworks proposed before.

\item caffe.Solver

Training a network involves a lot of tricks or policies to tune,
although these can be carefully designed and solidified at the backend.
We sometimes do require a flexible train procedure for a fine-grained training(e.g. critic mechanism introduced in \citet{Arjovsky2017}).
Thus, it is necessary to prepare the basics for warping Solver at the frontend.

We preset several immutable learning rate polices deployed in Caffe,
but implemented by a mutable learning rate which was common used in Theano codes(i.e. bind learning rate to a tensor).
Mutation can perform everywhere through workspace, which makes it simple to custom learning rate polices.

\end{itemize}

\subsection{TensorFlow}

Contrast to other graph based frameworks, TensorFlow evolves to be more engineering and scalable.
However, it still can be compounded of several existing designs we proposed above.
We attribute it to the efficiency of representation, concluded by \citet{Bengio2012},
that in a hierarchical deep architecture there is re-use of interfaces and sharing of sub-functions to build functions.

\begin{itemize}

\item tf.Session

\emph{Session} plays a similar role as \emph{Function}(\citet{Theano}). Fetches and feeds represent outputs and inputs individually.
What is different is that a session may consist of several functions,
and the pair of fetches and feeds is regarded as a unique key for hashing.

We propose a new structure called \emph{Transaction} to bind the auto-generated hashing key to a function.
Repeatly running session with an existing transaction will not lead to duplicated graphs,
which prefers to be an implicit way for creating and running graphs.

\item tf.placeholder \& tf.Variable

We do not distinguish them separately because Theano regards both as same in \emph{Function}.
Either \emph{placeholder} or \emph{Variable} requires a specific shape due to TensorFlow is a strongly shaped framework while Tiny-Dragon mentioned in Sec 4.2 is not.
Hence, we try to relax the shape-checking at the frontend for the following purposes:

\begin{itemize}

\item Most recent vision applications(\citet{Girshick2015}, \citet{Ren2015}, \citet{Shelhamer2017}, \citet{Gatys2015}, \citet{Johnson2016}) takes varied image shape.

\item Presetting shape for a learnable weight is tortured.

\item Feeding data with a arbitrary shape makes it unconstrained simulating diverse interfaces.

\end{itemize}

However, a restrict checking is also supported while inputs take valid shape.
We can use \emph{placeholder} or \emph{Variable} instead of \emph{Tensor} provided by Tiny-Dragon to declare inputs with premier shape,
and the following shape will be inferred throughout all available operators.
Note that the shapes inferred at the frontend do not totally equal to those inferred during runtime when taking a varied data shape.

\item tf.train.Optimizer

Recent popular frameworks, Caffe, MXNet, TensorFlow adopt coarse-gained optimizer unanimously,
which is a bit different from Theano proposed early.

Naively applying an update leads neither efficient computation nor convenience on large scale distributed training.
Sec 3.2 presents that we adopt an \emph{Updater} at the frontend for the optimization of objectives.
It accepts grads and variables to generate an independent graph for updating, which is similar to Optimizer in TensorFlow.
TensorFlow wraps computing and updating processes into a Train-Op to execute them together, which means that the Session
may take a group of solving objectives even if the fetching list only has one element.
Different from TensorFlow, we directly pass the Updater into a \emph{Function} to separate computing and updating process.
In order to combine both designs, we have no choice but record both gradients and respective objectives.

\end{itemize}

\bibliographystyle{iclr2017_conference}
\bibliography{iclr2017_conference}

\begin{thebibliography}{33}
\providecommand{\natexlab}[1]{#1}
\providecommand{\url}[1]{\texttt{#1}}
\expandafter\ifx\csname urlstyle\endcsname\relax
  \providecommand{\doi}[1]{doi: #1}\else
  \providecommand{\doi}{doi: \begingroup \urlstyle{rm}\Url}\fi

\bibitem[Abadi et~al.(2016)Abadi, Barham, Chen, Chen, Davis, Dean, Devin,
  Ghemawat, Irving, and Isard]{TensorFlow}
Mart¨ªn Abadi, Paul Barham, Jianmin Chen, Zhifeng Chen, Andy Davis, Jeffrey
  Dean, Matthieu Devin, Sanjay Ghemawat, Geoffrey Irving, and Michael Isard.
\newblock Tensorflow: a system for large-scale machine learning.
\newblock In \emph{Usenix Conference on Operating Systems Design and
  Implementation}, 2016.

\bibitem[Arjovsky et~al.(2017)Arjovsky, Chintala, and Bottou]{Arjovsky2017}
Martin Arjovsky, Soumith Chintala, and L¨¦on Bottou.
\newblock Wasserstein gan.
\newblock 2017.

\bibitem[Bengio(2009)]{Bengio2009}
Yoshua Bengio.
\newblock Learning deep architectures for ai.
\newblock \emph{Foundations \& Trends in Machine Learning}, 2\penalty0
  (1):\penalty0 1--55, 2009.

\bibitem[Bengio(2012)]{Bengio2012}
Yoshua Bengio.
\newblock Evolving culture vs local minima.
\newblock \emph{Computer Science}, 2012.

\bibitem[Bishop \& ChristopherM(2006)Bishop and ChristopherM]{Bishop2006}
Bishop and ChristopherM.
\newblock \emph{Pattern recognition and machine learning}.
\newblock Springer,, 2006.

\bibitem[Chen(2015)]{MXNet}
Tianqi Chen.
\newblock Mxnet: A flexible and efficient machine learning library for
  heterogeneous distributed systems.
\newblock \emph{Statistics}, 2015.

\bibitem[Chetlur et~al.(2014)Chetlur, Woolley, Vandermersch, Cohen, Tran,
  Catanzaro, and Shelhamer]{CUDNN}
Sharan Chetlur, Cliff Woolley, Philippe Vandermersch, Jonathan Cohen, John
  Tran, Bryan Catanzaro, and Evan Shelhamer.
\newblock cudnn: Efficient primitives for deep learning.
\newblock \emph{Computer Science}, 2014.

\bibitem[Cho et~al.(2014)Cho, Merrienboer, Gulcehre, Bahdanau, Bougares,
  Schwenk, and Bengio]{Cho2014}
Kyunghyun Cho, Bart~Van Merrienboer, Caglar Gulcehre, Dzmitry Bahdanau, Fethi
  Bougares, Holger Schwenk, and Yoshua Bengio.
\newblock Learning phrase representations using rnn encoder-decoder for
  statistical machine translation.
\newblock \emph{Computer Science}, 2014.

\bibitem[Elman(1990)]{Elman1990}
Jeffrey~L. Elman.
\newblock Finding structure in time ¡î.
\newblock \emph{Cognitive Science}, 14\penalty0 (2):\penalty0 179--211, 1990.

\bibitem[Gatys et~al.(2015)Gatys, Ecker, and Bethge]{Gatys2015}
Leon~A. Gatys, Alexander~S. Ecker, and Matthias Bethge.
\newblock A neural algorithm of artistic style.
\newblock \emph{Computer Science}, 2015.

\bibitem[Geman \& Geman(1984)Geman and Geman]{Geman1984}
S~Geman and D~Geman.
\newblock Stochastic relaxation, gibbs distributions, and the bayesian
  restoration of images.
\newblock \emph{IEEE Transactions on Pattern Analysis \& Machine Intelligence},
  PAMI-6\penalty0 (6):\penalty0 721--741, 1984.

\bibitem[Girshick(2015)]{Girshick2015}
Ross Girshick.
\newblock Fast r-cnn.
\newblock In \emph{IEEE International Conference on Computer Vision}, pp.\
  1440--1448, 2015.

\bibitem[Glorot et~al.(2011)Glorot, Bordes, and Bengio]{Glorot2011}
Xavier Glorot, Antoine Bordes, and Yoshua Bengio.
\newblock Deep sparse rectifier neural networks.
\newblock In \emph{International Conference on Artificial Intelligence and
  Statistics}, 2011.

\bibitem[Goodfellow et~al.(2014)Goodfellow, Pougetabadie, Mirza, Xu,
  Wardefarley, Ozair, Courville, and Bengio]{Goodfellow2014}
Ian~J. Goodfellow, Jean Pougetabadie, Mehdi Mirza, Bing Xu, David Wardefarley,
  Sherjil Ozair, Aaron Courville, and Yoshua Bengio.
\newblock Generative adversarial networks.
\newblock \emph{Advances in Neural Information Processing Systems}, 3:\penalty0
  2672--2680, 2014.

\bibitem[Griewank(2000)]{Griewank2000}
Andreas Griewank.
\newblock Evaluating derivatives.
\newblock \emph{Society for Industrial \& Applied Mathematics Philadelphia Pa},
  2000.

\bibitem[He et~al.(2015)He, Zhang, Ren, and Sun]{He2015}
Kaiming He, Xiangyu Zhang, Shaoqing Ren, and Jian Sun.
\newblock Deep residual learning for image recognition.
\newblock pp.\  770--778, 2015.

\bibitem[Hochreiter \& Schmidhuber(1997)Hochreiter and
  Schmidhuber]{Hochreiter1997}
S~Hochreiter and J~Schmidhuber.
\newblock Long short-term memory.
\newblock \emph{Neural Computation}, 9\penalty0 (8):\penalty0 1735--1780, 1997.

\bibitem[Jia et~al.(2014)Jia, Yangqing, Shelhamer, Evan, Donahue, Jeff,
  Karayev, Sergey, Long, and Jonathan]{Caffe}
Jia, Yangqing, Shelhamer, Evan, Donahue, Jeff, Karayev, Sergey, Long, and
  Jonathan.
\newblock Caffe: Convolutional architecture for fast feature embedding.
\newblock \emph{Eprint Arxiv}, pp.\  675--678, 2014.

\bibitem[Johnson et~al.(2016)Johnson, Alahi, and Li]{Johnson2016}
Justin Johnson, Alexandre Alahi, and Fei~Fei Li.
\newblock Perceptual losses for real-time style transfer and super-resolution.
\newblock 2016.

\bibitem[Jordan(1986)]{Jordan1986}
Michael~I Jordan.
\newblock Serial order: A parallel distributed processing approach.
\newblock 121:\penalty0 64, 1986.

\bibitem[Kingma \& Ba(2014)Kingma and Ba]{Kingma2014}
Diederik~P. Kingma and Jimmy Ba.
\newblock Adam: A method for stochastic optimization.
\newblock \emph{arXiv preprint arXiv:1412.6980}, 2014.

\bibitem[Krizhevsky et~al.(2012)Krizhevsky, Sutskever, and
  Hinton]{Krizhevsky2012}
Alex Krizhevsky, Ilya Sutskever, and Geoffrey~E Hinton.
\newblock Imagenet classification with deep convolutional neural networks.
\newblock In \emph{International Conference on Neural Information Processing
  Systems}, pp.\  1097--1105, 2012.

\bibitem[Lecun et~al.(1998)Lecun, Bottou, Orr, and M¨¹ller]{Lecun1998}
Yann Lecun, L¨¦on Bottou, Genevieve~B. Orr, and Klaus~Robert M¨¹ller.
\newblock Effiicient backprop.
\newblock \emph{Lecture Notes in Computer Science}, 1524\penalty0 (1):\penalty0
  9--50, 1998.

\bibitem[Lecun et~al.(2015)Lecun, Bengio, and Hinton]{Lecun2015}
Yann Lecun, Yoshua Bengio, and Geoffrey Hinton.
\newblock Deep learning.
\newblock \emph{Nature}, 521\penalty0 (7553):\penalty0 436--444, 2015.

\bibitem[Nair \& Hinton(2010)Nair and Hinton]{Nair2010}
Vinod Nair and Geoffrey~E. Hinton.
\newblock Rectified linear units improve restricted boltzmann machines.
\newblock In \emph{International Conference on Machine Learning}, pp.\
  807--814, 2010.

\bibitem[Ren et~al.(2015)Ren, He, Girshick, and Sun]{Ren2015}
S.~Ren, K.~He, R~Girshick, and J.~Sun.
\newblock Faster r-cnn: Towards real-time object detection with region proposal
  networks.
\newblock In \emph{International Conference on Neural Information Processing
  Systems}, pp.\  91--99, 2015.

\bibitem[Rumelhart et~al.(1986)Rumelhart, Hinton, and Williams]{Hinton1986}
David~E. Rumelhart, Geoffrey~E. Hinton, and Ronald~J. Williams.
\newblock \emph{Learning representations by back-propagating errors}.
\newblock MIT Press, 1986.

\bibitem[Shelhamer et~al.(2017)Shelhamer, Long, and Darrell]{Shelhamer2017}
Evan Shelhamer, Jonathon Long, and Trevor Darrell.
\newblock Fully convolutional networks for semantic segmentation.
\newblock \emph{IEEE Transactions on Pattern Analysis \& Machine Intelligence},
  PP\penalty0 (99):\penalty0 1--1, 2017.

\bibitem[Simonyan \& Zisserman(2014)Simonyan and Zisserman]{Simonyan2014}
Karen Simonyan and Andrew Zisserman.
\newblock Very deep convolutional networks for large-scale image recognition.
\newblock \emph{Computer Science}, 2014.

\bibitem[Srivastava et~al.(2014)Srivastava, Hinton, Krizhevsky, Sutskever, and
  Salakhutdinov]{Srivastava2014}
Nitish Srivastava, Geoffrey Hinton, Alex Krizhevsky, Ilya Sutskever, and Ruslan
  Salakhutdinov.
\newblock Dropout: a simple way to prevent neural networks from overfitting.
\newblock \emph{Journal of Machine Learning Research}, 15\penalty0
  (1):\penalty0 1929--1958, 2014.

\bibitem[Szegedy et~al.(2014)Szegedy, Liu, Jia, Sermanet, Reed, Anguelov,
  Erhan, Vanhoucke, and Rabinovich]{Szegedy2014}
Christian Szegedy, Wei Liu, Yangqing Jia, Pierre Sermanet, Scott Reed, Dragomir
  Anguelov, Dumitru Erhan, Vincent Vanhoucke, and Andrew Rabinovich.
\newblock Going deeper with convolutions.
\newblock pp.\  1--9, 2014.

\bibitem[{Theano Development Team}(2016)]{Theano}
{Theano Development Team}.
\newblock Theano: A python framework for fast computation of mathematical
  expressions.
\newblock \emph{arXiv e-prints}, abs/160Alex20125.02688, May 2016.
\newblock URL \url{http://arxiv.org/abs/1605.02688}.

\bibitem[Tieleman \& Hinton(2012)Tieleman and Hinton]{Tieleman2012}
T.~Tieleman and G.~Hinton.
\newblock Lecture 6.5-rmsprop: Divide the gradient by a running average of its
  recent manitude.
\newblock \emph{Technical Report: Neural networks for machine learning}, 2012.

\end{thebibliography}

\end{document}